# Reaction Coordinates are Optimal Channels of Energy Flow

Ao Ma* and Huiyu Li

Center for Bioinformatics and Quantitative Biology
Richard and Loan Hill Department of Biomedical Engineering
The University of Illinois at Chicago
851 South Morgan Street
Chicago, IL 60607

*correspondence should be addressed to:
Ao Ma
Email: aoma@uic.edu
Tel: (312) 996-7225

**Abstract**

Reaction coordinates (RCs) are the few essential coordinates of a protein that control its functional processes, such as allostery, enzymatic reaction, and conformational change. They are critical for understanding protein function and provide optimal enhanced sampling of protein conformational changes and states. Since the pioneering works in the late 1990s, identifying the correct and objectively provable RCs has been a central topic in molecular biophysics and chemical physics. This review summarizes the major advances in identifying RCs over the past 25 years, focusing on methods aimed at finding RCs that meet the rigorous committor criterion, widely accepted as the true RCs. Importantly, the newly developed physics-based energy flow theory and generalized work functional method provide a general and rigorous approach for identifying true RCs, revealing their physical nature as the optimal channels of energy flow in biomolecules.

# 1. Introduction

Proteins are the fundamental units of biological systems, responsible for most cellular functions. Over the past 50 years, intensive community efforts have substantially advanced our understanding of proteins, culminating in two monumental achievements. The advancements of the energy landscape theory settled the conceptual framework of protein folding (1-6); the development of AlphaFold solved the long-standing structure prediction problem (7; 8). With the native structures of proteins now readily available, the next major challenge in biophysics is to understand how protein conformational state changes and how these changes control protein function.

Starting with the pioneering work of Frauenfelder and colleagues on ligand rebinding to myoglobin in 1975 (9), a fascinating physical picture of the intricate relationship between protein conformation and protein function has gradually emerged (10-12). Proteins are dictated by a rugged energy landscape featuring many valleys corresponding to functionally important conformations, separated by barriers. Protein functions—such as enzymatic reactions, allostery, substrate binding, and protein-protein interactions—are governed by transitions between conformations. All these processes are thermally activated, driven by thermal fluctuations in the environment. Therefore, understanding protein function requires understanding activated processes.

## 1.1 Activated process.

An activated process requires the system to cross an activation barrier, which is significantly higher than the thermal energy $k_B T$, to transition from the reactant state to the product state. Consequently, the system needs to spend a prolonged waiting time in the reactant basin before it can cross the barrier, making this process orders of magnitude slower than the elementary molecular motions. This time-scale separation is a signature of activated processes, making them rare events. In 1889 (13), Arrhenius formulated the concept of activated processes from chemical reactions, which has since found many uses in diverse fields, such as chemical kinetics, diffusion in solids, homogeneous nucleation, and electrical transport, to name a few (14).

There are two central questions concerning activated processes: 1) the reaction rate, which is determined by how frequently the system crosses the activation barrier, and 2) the reaction mechanism, which concerns how the system acquires enough energy to overcome this barrier. Reaction rate theories, including transition state theory, RRKM theory, Kramers theory, Langer-Berezhkovskii-Szabo theory, and Grote-Hynes theory (14-32), provide both the conceptual framework for understanding these mechanisms and the practical procedures for calculating the rates. Although the reaction rate theories were derived from understandings of simple reactions in small molecules, the physical picture they present is so appealing that they have been widely adopted for proteins as well.

**1.2 Conceptual importance of reaction coordinates**.

All rate theories assume that a molecular system's activated dynamics, which are governed by the equations of motion (EoMs) of Newtonian mechanics, can be projected onto the motion of one or a few special coordinates, termed reaction coordinates (RCs), within a double-well potential. Different rate theories have developed distinct models of RC dynamics, which form the core of each theory and are based on different hypotheses or assumptions. Transition state theory assumes inertial dynamics at the barrier top, while Kramers theory assumes Langevin dynamics throughout the double-well, which is extended to a generalized Langevin equation in Grote-Hynes theory to account for memory effects. Thus, the concept of RC—an assumption rather than rigorous derivation from Newtonian mechanics—is the foundation of all reaction rate theories, placing it at the heart of the theoretical framework for understanding activated processes.

The concept of RC originated from chemical reactions of small molecules, which typically involve significant changes in only one degree of freedom (DoF), such as the formation of an H-H bond in $H + H_2$ (25; 33) or the rotation around a dihedral angle in butane (29). The RC is naturally identified with the DoF that accounts for all the changes in a reaction.

However, in a protein molecule that often contains tens of thousands of atoms or more, the existence of RCs is an extraordinarily bold hypothesis. The rationale is that the myriads of atoms in a protein, governed by Newtonian mechanics, can work together to perform functions in a chaotic cellular environment with high efficiency, selectivity, specificity, and tightly controlled timing, suggesting an intrinsic order and simplicity that reflect an underlying low-dimensional mechanism. The RCs in proteins, namely the few essential coordinates that fully control an activated process, embody this low-dimensional mechanism and encapsulate the essence of the activated dynamics, while all other coordinates (non-RCs) in the system are unimportant.

**1.3 Practical importance of RCs**.

RCs are also critically important for molecular dynamics (MD) simulations of biomolecules, which are indispensable for understanding how protein conformational states and changes control protein functions. MD simulations can provide all the details of protein dynamics at atomic resolution—an advantage unmatched by experimental techniques. However, their application is severely limited by two bottlenecks. First, the time scales easily accessible to MD simulations (a few hundred nanoseconds) are much shorter than the time scales of functionally important processes in proteins, which range from milliseconds to hours (10-12), making simulations infeasible except in a few special cases (34; 35). Second, the large number of DoFs in a protein makes it impossible to gain mechanistic insights from simulation data without effective dimension reduction. RCs provide optimal solutions to both bottleneck problems.

The main approach to addressing the first bottleneck is to enhance the sampling of protein conformational changes in MD simulations (36-38). This requires artificially accelerating activated processes, the only effective way to achieve these conformational changes in simulations. Many enhanced sampling methods, such as umbrella sampling (39; 40), metadynamics (38; 41; 42), adaptive biasing force (43; 44), apply bias potential on user-selected collective variables (CVs) (**sidebar: Collective Variables**) to accelerate activated processes. Their efficacy critically depends on the quality of the CVs: without suitable CVs, enhanced sampling provides no benefit over regular MD simulations (36).

It is well known that RCs are the optimal CVs for enhanced sampling (36; 37); the quality of CVs is determined by their overlap with the RCs. When CVs align with the RCs, the bias potential efficiently drives the RCs over the activation barrier, speeding up the activated process. In contrast, if CVs do not align with the RCs, the infamous 'hidden barrier' in the 'orthogonal space' will appear, preventing effective sampling (45; 46). The orthogonal space is the space orthogonal to the CVs, where the RCs lie; the hidden barrier is the actual activation barrier along the path of the RCs.

Efficacy of methods that guide system motion through selection and filtering, such as weighted ensemble simulation and non-equilibrium umbrella sampling (47-52), also relies on the overlap between RCs and user-selected CVs. These methods launch multiple independent MD trajectories, assess their progress along the CVs, and propagate or terminate them based on their advancements. If the CVs align with the RCs (53), the filtered trajectories will follow the natural transition pathway—the most efficient route for barrier crossing, thereby accelerating the activated process. (**sidebar callout: Natural transition pathway**) If not, trajectories will be misdirected to irrelevant high-energy regions in the conformational space and become trapped, rendering simulations slower than regular MD.

The second bottleneck can be overcome by developing a low-dimensional reduced description that captures the essence of the activated dynamics, allowing understanding the actual mechanisms and calculating reaction rates. If we can find the RCs, they provide the optimal reduced description. Unlike in small molecules, however, RCs in proteins are non-obvious. In practice, intuition-based ad hoc choices, such as root mean square deviation from reference structures, geometric parameters, and principal components, are widely used, although all without justification (54).

## 2. Rigorous Definition of RCs.

Given the fundamental importance of RCs and their hypothetical origin, a rigorous criterion for objectively validating the existence and correctness of RCs is indispensable. Pioneering works by Du et al. and Chandler's group in the late 1990s established this criterion based on the concept of the committor (55; 56). Only RCs that meet this criterion can fulfill their role in reaction rate theories and biomolecular simulations.

### 2.1 Committor.

Committor $p_B(R_0)$ is the probability that a dynamic trajectory initiated from a system conformation $R_0$, with initial momenta drawn from the Boltzmann distribution, reaches the product before the reactant state. The committor concept originated with Onsager's 1938 work on ion recombination (57), and was generalized by van Kampen in 1978 as the splitting probability for transitions between two stable states via Markovian processes (58). It was later used by Pratt and Ryter to define the transition state (TS) (59; 60). In the late 1990s to early 2000s, it was established as the general standard for understanding activated processes (55; 56).

Committor can be computed for any system conformation, requiring only the definitions of the reactant and product states. To compute $p_B(R_0)$, multiple MD trajectories (e.g., 50 to100 due to the high computational cost) are launched from $R_0$, each with different initial momenta. Committor is the ratio of trajectories that reach the product state to the total number of trajectories (55; 61; 62).

Since an activated process is a stochastic event, its probability of occurrence is the definitive variable for characterizing it. The committor $p_B(R_0)$ is the probability that an activated process will occur once the system reaches conformation $R_0$, making it the definitive variable for characterizing an activated process and quantifying its progress. It rigorously defines the important state of an activated process: reactant and product have committor values of 0 and 1, respectively, while the TS has $p_B = 0.5$. Transition path theory (63) demonstrates that all important dynamic properties of an activated process can be calculated using committor, including: 1) the probability of finding a reactive trajectory at $R_0$, 2) the flux of reactive trajectories, 3) the mean first passage time (64), 4) the mean transition path time (65), (**sidebar callout: Transition path time**) and 5) the reaction rate. Recently, Roux demonstrated that committor is optimal for calculating reaction rate in a double-well system (66).

## 2.2 Committor criterion for RCs.

Since committor encapsulates all the essential information of an activated process, the **true RCs** (tRCs) are defined as the few essential coordinates that fully determine the committor value of any system conformation (55; 56; 61; 62). All other coordinates (non-RCs), regardless of how many there are, do not affect the committor.

tRCs are the optimal CVs for enhanced sampling, provide the optimal reduced description for understanding activated processes, and ensure the correct transition rate in reaction rate theories (36; 55). In contrast, ad hoc RCs, which do not meet the committor criterion, are not useful for any of these purposes.

**2.3 Committor test for tRCs.**

Because committor can be computed for any system conformation without prior knowledge of tRCs, it provides an objective test for tRCs, known as the "committor test" (55; 56; 62; 67). In this test, an ensemble of test conformations is generated, all sharing the same values of the trial RCs corresponding to the TS but differing in other DoFs. Committor for each test conformation is then calculated. If the trial RCs are tRCs, committor values will be narrowly centered around 0.5. If the trial RCs are not tRCs, committor values will spread across the entire range from 0 to 1 or concentrate around 0 or 1 (Fig. 1). Therefore, passing the committor test is the necessary and sufficient condition for validating tRCs. As noted by Chandler and colleagues (55), calculating the mean of a committor distribution cannot validate tRCs, because even a uniform committor distribution $P(p_B) = 1$ has a mean of $\langle p_B \rangle = \int_0^1 P(p_B) p_B dp_B = 0.5$.

**3. Trial-and-error Approach for Identifying tRCs**

Identifying tRCs is a formidable task. Early works followed the conventional intuition-based trial-and-error approach. Du et al. applied the committor criterion to a Go model. They found that the number of native contacts, a common CV for protein folding, cannot determine the committor (56). Chandler and colleagues studied a range of processes (55; 68-70), demonstrating both the importance of the committor test and the formidable challenges in identifying tRCs.

**3.1 Lessons from alanine dipeptide.**

Notably, studies on alanine dipeptide by Chandler group have provided important clues for guiding subsequent investigations on tRCs in biomolecules. The $C_{7eq} \rightarrow C_{7ax}$ isomerization of alanine dipeptide in vacuum has become a benchmark system for developing methods for identifying tRCs (61; 62; 71-74). Although $C_{7eq}$ and $C_{7ax}$ are defined by the Ramachandran dihedrals $\phi$ and $\psi$, Bolhuis et al. found that the tRCs are $\phi$ and the dihedral $\theta_1$ (Fig. 2a), while $\psi$ is not a tRC (55). Why the counter-intuitive $\theta_1$ is a critical tRC had remained a puzzle for over 15 years (71; 72; 75; 76).

For the $C_{7eq} \rightarrow \alpha_R$ transition of alanine dipeptide in explicit water, Chandler and colleagues found that solute coordinates alone cannot determine $p_B$ because solvent molecules are a critical component of the tRC. Despite intense effort and innovative ideas, the solvent tRC eluded their investigations (68; 70).

**3.2 Challenges of identifying tRCs in complex molecules**

Early studies highlighted the formidable challenges of identifying tRCs in biomolecules. First, tRCs usually have counter-intuitive components, rendering human intuition inadequate and necessitating exhaustive explorations. This difficulty is compounded by the small number of tRCs compared to the overwhelmingly large number of DoFs in complex molecules, making identifying tRCs akin to finding needles in a haystack. Second, the high computational cost of committor calculation severely limits the number of trial RCs that can be tested; this number decreases sharply as system size increases. This limitation, along with the need for exhaustive searches, creates a deadlock for identifying tRCs using the trial-and-error approach, motivating the development of systematic methods to overcome these challenges.

**4. Machine Learning Methods for Identifying tRCs**

**4.1 The beginning**.

The first attempt of a systematic approach for identifying tRCs was to use machine learning (ML). The first ML method was the GNN method developed by Ma and Dinner in 2005 (62), which combines a genetic algorithm (GA) and a neural network (NN). The logic was that tRCs must closely correlate with committor since tRCs determine committor, making the committor a monotonic function of tRCs. The NN was used to learn this functional relationship.

This first example introduced three key elements of a generic ML method designed to address the challenges of identifying tRCs. First, it limits the cost of calculating committors to a fixed amount by employing a training set of system conformations (e.g. 2000) with a uniform distribution of pre-calculated $p_B$. Second, the NN uses a sigmoid function to model the relationship between the committor (i.e. the target variable) and a trial RC—a linear combination of user-provided CVs—and optimizes it by minimizing root mean square error (RMSE) between NN-predicted and precalculated committor values over the training set. This procedure compensates the inadequacy of human intuition. Third, the GA algorithm automates the search process by randomly selecting a combination of CVs from a user-provided pool and assigns their coefficients in the trial RC.

Using the GNN method, Ma and Dinner successfully identified tRCs for both the $C_{7eq} \to C_{7ax}$ and $C_{7eq} \to \alpha_R$ transitions of alanine dipeptide (62). Remarkably, they discovered the critical solvent tRC for the $C_{7eq} \to \alpha_R$ transition in water, which had eluded previous investigators (68; 70). This tRC is a torque that imposes an overall twisting on the solute towards the $\alpha_R$ state, arising from the electrostatic interactions from all the water molecules in the simulation box. This tRC is counter-intuitive yet physically meaningful, demonstrating the advantages of ML over human intuition. In an intriguing development, Lu et al. achieved substantial acceleration of the conformational transitions of deca-alanine by separately enhancing compensating fluctuations in solute-solute and solute-solvent interactions (77). Their results suggest that solvent fluctuations driving solute conformational transitions might be a general mechanism in biomolecular systems.

The success of the GNN method inspired further development of ML methods for identifying tRCs, a trend significantly boosted by the recent surge in ML applications. These methods differ in their adaption of the three key elements discussed above, leading to variations in their effectiveness at identifying tRCs. They fall into two categories: 1) methods based on the committor-tRC correlation (73; 78-82), and 2) methods assuming that tRCs maximally reconstitute the dynamics of the full system (74; 83). Additionally, an important class of related methods is based on the long-standing but somewhat vague notion that tRCs are the slowest DoFs in a system (84-86).

**4.2 Methods based on committor-tRC correlation.**

An early example is the maximum likelihood method by Peters and Trout (78; 87-91). Similar to the GNN method, it used a sigmoid to fit the committor as a function of the trial RC, which is a linear combination of user-provided CVs. They developed the aimless shooting algorithm to reduce the cost of calculating $p_B$, albeit at the expense of reduced accuracy. This algorithm results in a $p_B$ distribution concentrated around the TS for the training data. Instead of RMSE, they maximized the likelihood function, $L = \prod_{p_B(R_0)} p_B(r(R_0)) \prod_{p_B(R_0)} \left(1 - p_B(r(R_0))\right)$, to optimize the trial RC. Here, $p_B(R_0)$ and $p_B(r(R_0))$ are the committor value of $R_0$ by aimless shooting and the sigmoid function, respectively. Later, Lechner et al. introduced non-linearity by replacing the linear combination of CVs with a string of conformations in a low-dimensional CV space (87).

The method by Schwartz and colleagues took a novel perspective. They focused on the behaviors of molecular coordinates at the TS (79; 92), assuming that the tRCs should show the least variation because the stochastic separatrix spanned by TS conformations should be 'thin' along the direction of the tRC. They used principal component analysis (PCA) to identify coordinates with the least variation. To account for the nonlinear relationship between the tRCs and molecular coordinates, they later introduced the kernel PCA method (93-95).

Recently, Jung et al. combined features of GNN and maximum likelihood methods (81; 96). The $p_B$ distribution over their training data is concentrated around the TS; the trial RC is optimized by maximizing the likelihood function. In the method by Mori and Saito (82), the authors prepared the training data by randomly extracting system conformations from transition path sampling (TPS) trajectories, resulting in an uncertain $p_B$ distribution over the training set. (**sidebar callout: Transition Path Sampling**) They optimized the trial RC by minimizing the cross entropy between predicted and computed committor values.

Frassek et al. used an extended autoencoder for identifying RCs (80). They introduced a committor decoder to learn the relationship between the committor and the trial RC, which is a linear combination of user-provided CVs. To train the decoder, they binned the CVs and estimated $p_B$ for each bin as the fraction of reactive trajectories passing through that bin. Instead of the committor test, they used correlation between predicted and estimated committor values to evaluate the quality of optimized trial RCs.

Mori et al. used a deep NN to identify RCs in the benchmark system of $C_{7eq} \rightarrow C_{7ax}$ isomerization of alanine dipeptide in a vacuum (73; 97). They represented the trial RC using the deep NN and optimized it by minimizing the cross entropy between predicted and computed committors. Like the GNN method, their training set features a uniform $p_B$ distribution. Notably, they identified the $\theta_1$ dihedral as a tRC, a target missed by other post-GNN ML methods.

**4.3 Methods based on reconstituting dynamic information**.

These methods aim to identify low-dimensional CVs that can best preserve the dynamics of the full system. One example is the reweighted autoencoded variational Bayes for enhanced sampling (RAVE) by Tiwary and colleagues (83; 98-100). They used a linear encoder to generate the trial RC from user-provided CVs and optimized it by minimizing the cross entropy between encoder and decoder distributions. Bias is then applied to trial RCs to generate new trajectories, which are used to refine trial RCs iteratively. Notably, in the application of RAVE to alanine dipeptide in vacuum, the CV pool did not include $\theta_1$, so it remains to be seen whether RAVE can recognize $\theta_1$ as a tRC.

The method by Ensing and colleagues (74) utilized a combination of a GA and an NN, but the scoring function is the molecular conformations from TPS trajectories rather than their committor values. They applied this method to the benchmark $C_{7eq} \to C_{7ax}$ isomerization of alanine dipeptide but did not identify $\theta_1$ as a tRC. The possible reason is that conformations on TPS trajectories contain many features unrelated to the $C_{7eq} \to C_{7ax}$ transition and committor, causing the NN to assign low significance to $\theta_1$.

**4.4 Methods focusing on the slow modes**.

These methods were motivated by the assumption that tRCs are the slowest DoFs in an activated process, as they determine the transition rate. For a system governed by Markovian dynamics, all time scales can be obtained from spectral decomposition of the transition operator, which cannot be directly computed. Consequently, Noe and colleagues proposed the variational approach to conformational dynamics (VAC) (84; 85). This approach finds the eigenfunctions of the transition operator by solving the eigenvalue problem of the time-lagged covariance matrix of the basis functions used for eigenfunction expansion.

VAC-based methods have two key elements: 1) basis functions, and 2) the lag time. Different choices of these elements lead to different variants with varying performances (101). The Markov state model (MSM) uses the indicator functions of metastable states as the basis functions (102-105), while the time-lagged independent component analysis (TICA) uses linear basis functions (106-109). The variational approach for Markov processes (VAMP) generalizes VAC to treat non-reversible processes (110), where the time-lagged covariance matrices become asymmetric and need to be solved by singular value decomposition. The VAMPnet method (86) uses NN to optimize basis functions.

Interestingly, in the application of VAMPnet to alanine dipeptide in vacuum, the leading eigenfunction does not contain the critical $\theta_1$ dihedral required to meet the committor criterion. This result suggests the difference between tRCs and the slow modes identified by VAC methods. One possible reason is the existence of slow dynamics uncoupled or orthogonal to the actual activated process, a well-known caveat with VAC methods. Another caveat of VAC methods is the choice of the lag time. While the results are sensitive to the lag time, there is no rigorous principle for how to choose it (111; 112).

**4.5 Caveats of ML in the identification of tRCs**.

Despite the success of ML in identifying tRCs, there are daunting hurdles. ML excels at recognizing sophisticated patterns by building a model of the pattern embedded in the target variables (e.g. committor or dynamics of the full system) using user-provided CVs. The pattern discovered by ML is represented by the model optimized by the algorithm, and its quality is assessed by its ability to reproduce target variables using metrics such as RMSE or cross entropy.

The critical pattern for identifying tRCs is the tRC dynamics during barrier crossing. Outside the narrow window of the transition period, tRC dynamics within the stable basins are no different from that of non-RC dynamics. Therefore, a successful ML method must meet two necessary conditions: (1) the target variables need to accurately capture the barrier-crossing dynamics of tRCs, and (2) the user-provided CVs need to provide sufficient information for the ML algorithm to build a model that can reproduce the target variables.

Without *a priori* knowledge of tRCs, committor is the only target variable that reliably satisfy the first condition, because barrier crossing is transient and tRCs are vastly outnumbered by non-RCs. Without the committor to distill barrier-crossing dynamics of tRCs from the overall system dynamics, tRCs are like a needle in the haystack within the enormous conformational space dominated by non-RCs. When the target variables include a significant amount of non-committor information, ML models will prioritize reproducing this information, as it overwhelmingly outweighs committor information, leading to most likely incorrect results. This is why, among all the post-GNN ML methods applied to the $C_{7eq} \rightarrow C_{7ax}$ isomerization of alanine dipeptide in vacuum, only the method by Mori et al. successfully identified the critical $\theta_1$ dihedral (73; 97). Their success is due to using accurate committor information and a rigorous weighting scheme, provided by the uniform committor distribution introduced in the GNN method.

To meet the second condition, the pool of user-provided CVs must include all the important tRC components. This is the reason that Ma and Dinner had to search through more than 6,000 CVs to find the solvent tRC for the $C_{7eq} \rightarrow \alpha_R$ transition of alanine dipeptide (62). This process is largely manual and intuition-based, making it extremely challenging for larger proteins.

Most importantly, tRCs are an emergent physical property of protein molecules, governed by underlying physical laws. Understanding these laws is far more valuable for understanding protein functions than merely identifying tRCs. Although current ML methods exceled at uncovering hidden patterns, they struggle to uncover these physical laws. For example, while AlphaFold has revolutionized structural biology by accurately predicting protein structure, or recognizing the 3D structural patterns in protein sequences (7; 8), it offers little insight into the mechanism of protein folding (113)—the physical principles that determine protein structure.

**5. Methods based on reaction rate theories**.

Since activated processes are governed by physical laws and tRCs are intrinsic physical properties of molecular systems, it is natural to identify tRCs based on physical principles. The key question is: what physical principle determines tRCs? In this regard, there are two major directions.

The first direction is built on the conceptual framework and mathematical formulation of reaction rate theories. These methods fall into three main categories: 1) minimizing recrossing in the TS region, 2) the pathway of maximum reactive flux, and 3) the minimum energy or free energy pathway.

**5.1 Methods on minimizing recrossing**

The earliest attempt in this category is the variational transition state theory (114; 115), which deems the direction of the minimum reactive flux in the saddle region as the RC. This direction minimizes the recrossing of dynamic trajectories launched from the TS and maximizes the transmission coefficient in transition state theory. Along similar lines, Pietrucci and colleagues proposed to identify optimal trial RC by minimizing the rate $k_{AB} = (\tau_{AB} + \tau_{BA})^{-1}$ for transitions between two states, A and B (116). Here, $\tau_{AB}$ and $\tau_{BA}$ are the mean first passage time for $A \rightarrow B$ and $B \rightarrow A$ transitions, respectively.

## 5.2 Methods on maximizing reactive flux

Berkowitz et al. derived a rigorous formalism for the path with maximum reactive flux for a diffusion-controlled reaction governed by the Smoluchowski equation (117). This formalism was later developed into a numerical algorithm by Huo and Straub (118). More recently, Elber and colleagues developed a method along similar lines using directional milestoning (119). They assume RC as the sequence of jumps between milestones connecting two metastable states that has the maximum flux based on transition rates computed from milestoning.

## 5.3 Methods on minimum energy path

The first example in this category is the method by Elber and Karplus (120), who defined the minimum energy path connecting two conformations as the path that minimizes a path functional built on the potential energy of the system. Two prominent examples are the nudged elastic band method (121-123) and the string method (124-129). Both consider a curve connecting two stable states in a low-dimensional space spanned by user-selected CVs, consisting of a chain of intermediate states known as images. Both methods move the trial curve towards the minimum energy path using the force perpendicular to the curve. To prevent the force tangent to the curve from collapsing all the images into the two basins, the nudged elastic band method introduces a spring between neighboring images to maintain proper spacing, while the string method uses a Lagrange multiplier to impose an equal arc length between neighboring images.

There are two important considerations with these methods. First, the user-selected CVs need to capture the essence of the activated process, which requires significant overlap with the tRCs—an inherently challenging task. Second, the optimization procedure is local in nature, making it prone to being trapped in local minima. This issue becomes more severe with increasing dimensionality and the ruggedness of the underlying energy landscape. For instance, in an application to the benchmark $C_{7eq} \rightarrow C_{7ax}$ transition of alanine dipeptide (130), the string method did not recognize dihedral $\theta_1$ as a tRC. While the swarm of trajectories method aims to alleviate the issue of becoming trapped in local minima (131), it operates on the assumption that short trajectories can surmount barriers, which often does not hold.

**6. Energy Flow Theory and Generalized Work Functional.**

Methods based on reaction rate theories rely on reduced descriptions and pre-assumed models of activated dynamics. However, these assumed low-dimensional dynamic models may not capture all the essential aspects of the full dynamics of naturally occurring transitions and invariably lead to challenging high-dimensional optimization problems. In contrast, the foundational first principle is that Newton's laws govern the behavior of all molecules, as quantum effects are negligible unless the formation or breaking of chemical bonds is involved. This principle is the foundation for the second direction of physics-based methods for identifying tRCs, which include the energy flow theory (EFT) (71; 75; 76) and the generalized work functional (GWF) method (72; 132; 133). They directly compute tRCs from the full system dynamics without involving any optimization problem.

**6.1 Energy is the cost function of motion.**

In the Lagrange-Hamiltonian formulation of Newton's laws, energy is the generating function of the equations of motion, making it the exact cost function of motion. tRCs can be identified as the coordinates that incur the highest energy cost because they must overcome the activation barrier. However, the potential energy function is dominated by complex couplings between different coordinates, making it impossible to rigorously define the energy of each coordinate and calculate the cost of its motion based on changes in its energy.

While it is possible to define energy per coordinate based on intuition, this involves making assumptions for partitioning energy among coordinates that go beyond Newton's laws, effectively violating them. Consequently, such an approach is counterproductive because it destroys the exactness and rigor of energy as a cost function.

Fortunately, the fundamental principles of multivariable calculus enable us to directly define the exact energy cost per coordinate. The total differential of a generic *n*-variable function $f(x_1, \ldots, x_n)$ is:

$$df = \sum_{i=1}^{n} \frac{\partial f}{\partial x_i} dx_i \quad (1).$$

Here, each partial differential $\frac{\partial f}{\partial x_i} dx_i$ is the exact contribution of the change in $x_i$ to the total change in $f(x_1, \ldots, x_n)$. The partial derivatives $\frac{\partial f}{\partial x_i}$ exactly partition the coupling terms of $f(x_1, \ldots, x_n)$ among all the variables, regardless of their complexity. A finite change $\Delta f = f_1 - f_0$ can be obtained by integrating the differential change:

$$\Delta f = \int_{f_0}^{f_1} df = \sum_{i=1}^{n} \int_{x_{i,0}}^{x_{i,1}} \frac{\partial f}{\partial x_i} dx_i \quad (2).$$

Consequently, $\int_{x_{i,0}}^{x_{i,1}} \frac{\partial f}{\partial x_i} dx_i$ defines the exact contribution of changes in $x_i$ (i.e. $x_{i,1} - x_{i,0}$) to the accumulated change in $f(x_1, \ldots, x_n)$. Applying this basic concept to the potential energy $U(\boldsymbol{q})$ and the kinetic energy $K(\boldsymbol{q}, \dot{\boldsymbol{q}}) = \frac{1}{2}\sum_i^N p_i \dot{q}_i = \frac{1}{2}\sum_{i=1}^N s_{ij}\dot{q}_i\dot{q}_j$ leads to rigorously defined potential and kinetic energy cost of each coordinate $q_i$ that are exact, without any assumption or approximation. They are called the potential and kinetic energy flow of each coordinate, respectively. Here, $s_{ij} = \sum_{\alpha=1}^{N} m_\alpha \frac{\partial x_\alpha}{\partial q_i} \frac{\partial x_\alpha}{\partial q_j}$ is the structural coupling factor between coordinates $q_i$ and $q_j$; $x_\alpha$ is a Cartesian coordinate and $m_\alpha$ the mass of the corresponding atom.

### 6.2 Potential energy flow.

The exact potential energy flow (PEF) through a coordinate $q_i$, in its differential form $dW_i$ and finite form $\Delta W_i(t_1, t_2)$, is the mechanical work performed on $q_i$ (71):

$$dW_i = -\frac{\partial U(\boldsymbol{q})}{\partial q_i} dq_i = -\frac{\partial U(\boldsymbol{q})}{\partial q_i} \dot{q}_i dt;$$

$$\Delta W_i(t_1, t_2) = \int_{t_1}^{t_2} dW_i = -\int_{t_1}^{t_2} \frac{\partial U(\boldsymbol{q})}{\partial q_i} \dot{q}_i dt \quad (3).$$

Here, $\Delta W_i(t_1, t_2)$ is the change in $U(\boldsymbol{q})$ caused by the motion of $q_i$ alone. It projects the change in the total potential energy onto the motion of $q_i$ and quantifies its cost. In a complete and orthogonal coordinate system, $dU(\boldsymbol{q}) = -\sum_{i=1}^{N} dW_i$ exactly partitions the change in $U(\boldsymbol{q})$ among all the coordinates, ensuring no overlap between the PEFs of any two coordinates.

### 6.3 $dW_i$ is the generating function of the EoM of an individual coordinate.

The heart of the Hamiltonian equations is $\dot{p}_i = -\frac{\partial H}{\partial q_i} = -\frac{\partial K}{\partial q_i} - \frac{\partial U}{\partial q_i}$. Moving $\frac{\partial K}{\partial q_i}$ to the left-hand side of the equation results in $\dot{p}_i + \frac{\partial K}{\partial q_i} = -\frac{\partial U}{\partial q_i}$, and Eq. (3) gives $\frac{dW_i}{dq_i} = -\frac{\partial U}{\partial q_i}$. Combining these, we have:

$$\dot{p}_i + \frac{\partial K}{\partial q_i} = \frac{dW_i}{dq_i} \quad (4)$$

The left-hand side of Eq. (4) represents the change in the motion of $q_i$, thus the EoM that determines the dynamics of $q_i$ is generated by taking derivative of $dW_i$, making $dW_i$ the generating function of this EoM. Consequently, the PEF vector $d\boldsymbol{W} = (dW_1, \ldots, dW_N)$ encompasses the complete information of the dynamics of each coordinate in the system.

### 6.4 Kinetic energy flow.

The exact kinetic energy flow (KEF) through $q_i$, in its differential form $dK_i$ and finite form $\Delta K_i(t_1, t_2)$, is defined as (76):

$$dK_i = \frac{\partial K}{\partial \dot{q}_i} d\dot{q}_i + \frac{\partial K}{\partial q_i} dq_i = \left( p_i \ddot{q}_i + \frac{\partial K}{\partial q_i} \dot{q}_i \right) dt;$$

$$\Delta K_i(t_1, t_2) = \int_{t_1}^{t_2} dK_i = \int_{t_1}^{t_2} \left( p_i \ddot{q}_i + \frac{\partial K}{\partial q_i} \dot{q}_i \right) dt \quad (5).$$

Here, $\Delta K_i(t_1, t_2)$ is the change in $K(\boldsymbol{q}, \dot{\boldsymbol{q}})$ caused by the motion of $q_i$ alone and a measure of its cost. Accordingly, $dK(\boldsymbol{q}, \dot{\boldsymbol{q}}) = \sum_{i=1}^{N} dK_i$ provides the exact partition of a change in $K(\boldsymbol{q}, \dot{\boldsymbol{q}})$ among all the coordinates, ensuring no overlap between the KEFs of any two coordinates.

### 6.5 Applications of EFT.

Energy flow theory provides a general and exact approach for analyzing the mechanisms of dynamic processes, with two additional ingredients: a suitable coordinate system and a proper ensemble average.

For biomolecules, internal coordinates of bonds, angles, and dihedrals are the proper coordinate system because they provide a natural description of protein motion that automatically satisfies all constraints from bonded interaction. In contrast, movements of Cartesian coordinates are dominated by structural restraints from bonded forces, resulting in a convoluted physical picture.

The ensemble average of an energy flow $A(\Gamma)$ in phase space $\Gamma$, in its differential form $\langle dA(\xi^*) \rangle$ and finite form $\langle \Delta A(\xi_1, \xi_2) \rangle$, is defined by:

$$\langle dA(\xi^*) \rangle = \frac{\int \rho(\Gamma) \delta(\xi(\Gamma) - \xi^*) dA[\xi(\Gamma) \to \xi(\Gamma) + d\xi] d\Gamma}{\int \rho(\Gamma) \delta(\xi(\Gamma) - \xi^*) d\Gamma};$$

$$\langle \Delta A(\xi_1, \xi_2) \rangle = \int_{\xi_1}^{\xi_2} \langle dA(\xi) \rangle \quad (6).$$

Here, $\xi(\Gamma)$ is a projector variable that quantifies the progress of the target dynamic process. For an activated process, $\xi(\Gamma)$ can be the committor or a CV that clearly distinguishes reactant and product states (71; 72; 133). For an energy relaxation process, time itself is a good projector (132; 134). $\rho(\Gamma) d\Gamma$ represents the probability of finding the system around a phase-space point $\Gamma$ in the ensemble of dynamic trajectories of the target process. $\delta(\xi(\Gamma) - \xi^*)$ is the Dirac $\delta$-function; $\xi^*$ is a specific value of $\xi(\Gamma)$. $dA[\xi(\Gamma) \to \xi(\Gamma) + d\xi]$ projects the change in $A(\Gamma)$ along each trajectory in the ensemble onto the differential interval at $\xi(\Gamma)$; it is the change in $A(\Gamma)$ over the interval $[\xi(\Gamma), \xi(\Gamma) + d\xi)$. $\Delta A(\xi_1, \xi_2)$ is the change in $A$ over the finite interval $[\xi_1, \xi_2]$.

**6.6 Mechanism of activated processes: Kramers picture**.

The main purpose for studying activated processes is to understand their underlying mechanisms. Kramers theory depicts the standard physical picture of an activated process, consisting of two critical steps (Fig. 3a): energy activation (EA) and barrier crossing (BC). EA is when the tRC gathers sufficient energy to overcome the activation barrier, and BC is the actual crossing of this barrier. BC cannot occur without EA.

In the Kramers picture, EA and BC proceed hand-in-hand as the system climbs the activation barrier. During this process, the total energy of the tRC increases stochastically, driven by random forces from the thermal bath, and peaks when the system reaches the barrier top.

Kramers theory is foundational for understanding activated processes in biomolecules, commonly used to interpret experiments on enzymatic reactions, conformational changes, and protein folding. However, Kramers theory was developed for small molecule reactions in solution, while proteins are complex molecules. Small molecules require an external thermal bath to provide activation energy, whereas complex molecules, including alanine dipeptide, have enough DoFs to form an intramolecular thermal bath, supplying the tRCs with the necessary energy. Unlike an external bath, which couples to tRCs only through non-bonded interactions, an intramolecular bath couples to tRCs via both bonded and non-bonded interactions, resulting in a fundamentally different interplay between the tRCs and the thermal bath.

Consequently, Kramers theory faced significant challenges when applied to proteins, becoming more evident with advances in experimental techniques that enabled more detailed measurements. For example, Neupane et al. found inconsistencies between the barrier height and diffusion constant extracted from single-molecule protein folding data and the measured transition path time (135-138).

**6.7 Mechanism of activated processes: energy flow picture**.
Therefore, it is essential to rigorously analyze activated processes in complex molecules and compare the results with the Kramers picture. Energy flow theory is particularly well-suited for this purpose as it quantifies the exact energy change of each coordinate, allowing for direct comparison to the Kramers picture, which centers on the total energy of the tRC. The $C_{7eq} \rightarrow C_{7ax}$ transition of alanine dipeptide, the smallest complex molecule, is an ideal test system. Application of EFT to this process (75; 76; 139) revealed an intriguing mechanism fundamentally different from the Kramers picture.

**6.7.1 Coordination between EA and BC**. Contrary to the Kramers picture, energy flow analyses revealed that EA and BC proceed sequentially, with EA preceding BC and lasting five times longer (Fig. 3b) (75). During EA, potential energy undergoes no systematic change, while kinetic energy flows from the non-RCs into $\phi$, accumulating until it matches the activation barrier height (Fig. 3c). At this point, EA ends and BC begins. During BC (71), $\phi$ moves against the forces exerted by the non-RCs, converting its kinetic energy accumulated during EA into potential energy as it climbs the activation barrier (Fig. 3c). This conversion continues until $\phi$ reaches the barrier top at $p_B = 0.5$. Afterwards, $\phi$ relaxes into the $C_{7ax}$ basin, and the potential energy converts back into kinetic energy (Fig. 3c).

Strikingly, during BC, $\theta_1$ receives potential energy from the non-RCs and uses it to assist $\phi$ in climbing the activation barrier (Fig. 3c, d). This tight coordination between $\theta_1$ and $\phi$ explains why $\theta_1$ is a tRC and results in vortex-like rotational fluxes in the TS region—a surprising feature distinct from the diffusion-dominated Kramers picture (140).

**6.7.2 Direct transfer and accumulation of kinetic energy.** Kinetic energy flows into $\phi$ and accumulates during EA through a novel mechanism: it directly transfers from non-RCs into $\phi$ without non-RCs doing work on $\phi$ (75). The curvilinear nature of internal coordinates enables kinetic energy to transfer from one coordinate $q_i$ to another coordinate $q_j$ via the structural coupling $s_{ij}$ between them, requiring only proper coherence between the dynamics of $q_i$ and $q_j$ (76). This mechanism contrasts with the conventional understanding based on Cartesian coordinates, where a coordinate's kinetic energy increases only when external forces are doing work on it to persistently increase its velocity.

While a fascinating mechanism, a critical question remains: Can it be effective in solution? The concern is that random collisions with solvent molecules will change the velocity of the tRC chaotically and prevent its sustained increase, which is required for accumulating kinetic energy in Cartesian coordinates.

The answer is yes. In internal coordinates, kinetic energy can transfer to and accumulate in tRCs even under chaotic collisions with solvents. Kinetic energy accumulates in $\phi$ through positive KEF into $\phi$, defined as $\langle dK_\phi \rangle = \left( \frac{\partial K}{\partial \phi} \dot{\phi} + \frac{\partial K}{\partial \dot{\phi}} \ddot{\phi} \right) dt > 0$, based on Eqs. (5) and (6). This process does not require $\ddot{\phi} > 0$ as it does in Cartesian coordinates. Instead, positive correlations between $\frac{\partial K}{\partial \phi}$ and $\dot{\phi}$, as well as between $\frac{\partial K}{\partial \phi}$ and $\ddot{\phi}$, are sufficient. Since both $\frac{\partial K}{\partial \phi}$ and $\frac{\partial K}{\partial \phi}$ are functions of protein structure, proper structural fluctuations, sustained even under constant solvent collisions, can ensure persistent correlations and the accumulation of kinetic energy in $\phi$.

**6.7.3 The energy flow picture**. The findings above revealed an intriguing physical picture (71; 75). The movements of individual coordinates are driven by the energy flowing through them, making each coordinate an energy flow channel. While the coordinates move chaotically, their energy flows exhibit order, stemming from each channel's capacity determined by the protein structure and fine-tuned through evolution. During EA and BC, energy—either kinetic or potential—flows systematically from non-RCs into tRCs, assisting them in crossing the activation barrier. Consequently, substantial energy flows through the tRCs, underscoring their physical nature as the optimal channels with the highest capacities.

**7. Generalized Work Functional**.

The tight cooperativity between $\theta_1$ and $\phi$ (Figs. 3c, d) suggests that each coordinate has a major component along the tRC, as well as a minor component orthogonal to it, indicating that tRCs and non-RCs are entangled in internal coordinates. This entanglement causes poor time-scale separation between different coordinates, leading to a rugged free energy landscape and a convoluted energy flow pattern (Fig. 4).

Therefore, the key to identifying tRCs is to find the 'ideal' coordinate system that cleanly separates tRCs from non-RCs. This will maximize time-scale separation between tRCs and non-RCs, leading to a smoothed free energy landscape and systematic energy flows (Fig. 4). Since the PEF of each coordinate is the generating function (Eq. (4)) of its EoM and its most important feature, the "ideal" coordinate system should maximize the differences between the PEFs of different coordinates. This can be achieved through proper transformation of the PEF vector $d\mathbf{W} = (dW_1, \ldots, dW_N)$ between different coordinate systems.

**7.1 Coordinate transformation of $d\mathbf{W}$ requires generalizing the concept**.

The transformation of $dW_i = F_i dq_i$ in coordinates $\boldsymbol{q}$ to $dW_\alpha = F_\alpha dr_\alpha$ in coordinates $\boldsymbol{r}$, related by $d\boldsymbol{r} = \boldsymbol{A} \cdot d\boldsymbol{q}$ and $d\boldsymbol{q} = \boldsymbol{A}^{-1} \cdot d\boldsymbol{r}$, is given by the chain rule:

$$dW_\alpha = -\frac{\partial U}{\partial r_\alpha} dr_\alpha = -\sum_{i,k=1}^{N} \frac{\partial U}{\partial q_i} \frac{\partial q_i}{\partial r_\alpha} \frac{\partial r_\alpha}{\partial q_k} dq_k = \sum_{i,k=1}^{N} A_{\alpha i}(F_i dq_k) A_{k\alpha}^{-1} \quad (7).$$

Here, $A_{\alpha i} = A_{i\alpha}^{-1} = \frac{\partial r_\alpha}{\partial q_i} = \frac{\partial q_i}{\partial r_\alpha}$ because $\boldsymbol{A}$ is an orthogonal matrix. Equation (7) introduces a conceptually new physical quantity $F_i dq_k$, which is not a mechanical work. Therefore, the coordinate transformation of $d\boldsymbol{W}$ requires generalizing the concept of mechanical work to incorporate quantities like $F_i dq_k$.

Coordinate transformation is essentially a shift in perspective; representing $dW_\alpha$ in $\boldsymbol{q}$ requires projecting both $F_\alpha$ and $dr_\alpha$ onto all coordinates in $\boldsymbol{q}$. This results in the projected components $F_i dq_k$. Since $dW_\alpha$ represents the impact of the force $F_\alpha$ on the coordinate $r_\alpha$, $F_i dq_k$ represents the impact of $F_i$ on $q_k$. Equation (7) shows that $dW_\alpha$ is obtained by summing over all the components $F_i dq_k$, each weighted by the projection coefficients $\frac{\partial r_\alpha}{\partial q_i}$ and $\frac{\partial r_\alpha}{\partial q_k}$.

**7.2 GWF transforms $d\boldsymbol{W}$ between coordinate systems**.

The generalized work functional (GWF) generalizes the concept of mechanical work. Its differential in coordinate system $\boldsymbol{q}$ is defined as:

$$d\mathbb{W}_q = \boldsymbol{F} \otimes d\boldsymbol{q} \quad (8).$$

Here, $\otimes$ denotes tensor product, making $d\mathbb{W}_q$ an asymmetric tensor and $F_i dq_k$ its elements. Consequently, $d\mathbb{W}_q$ encompasses the comprehensive information of the impacts of all forces on the system's dynamics. The GWF in coordinates $\boldsymbol{r}$ and $\boldsymbol{q}$ are related by a similarity transformation:

$$d\mathbb{W}_r = \boldsymbol{A} \cdot d\mathbb{W}_q \cdot \boldsymbol{A}^{-1} \quad (9).$$

Therefore, the coordinate transformation of GWF does not introduce any new quantities, making it a self-contained fundamental concept with mechanical work as its sub-concept. All the important mechanical quantities, such as $d\boldsymbol{W}$ and $dU$, are encompassed in GWF:

$$d\boldsymbol{W}_r = \text{diag}(d\mathbb{W}_r) = \text{diag}(\boldsymbol{A} \cdot d\mathbb{W}_q \cdot \boldsymbol{A}^{-1}) \quad (10)$$

$$dU = \text{Tr}(d\mathbb{W}_r) = \text{Tr}(d\mathbb{W}_q) \quad (11),$$

where diag(·) denotes diagonal vector and Tr(·) denotes trace. Equation (10) further shows that $d\mathbb{W}$ is the operator for transforming $dW$ between coordinate systems.

### 7.3 Left singular vectors of GWF constitute the 'ideal' coordinate system.

The goal of the "ideal" coordinate system is to maximize the differences between the PEFs of different coordinates. To achieve this goal, we need to find an orthogonal coordinate system that maximizes the PEF of each coordinate. This can be achieved through the singular value decomposition of GWF:

$$d\mathbb{W} = \boldsymbol{U} \cdot \boldsymbol{\Lambda} \cdot \boldsymbol{V}^T \quad (12),$$

which decomposes $d\mathbb{W}$ in terms of its optimal basis tensors $\boldsymbol{u}_i \otimes \boldsymbol{v}_i$: $d\mathbb{W} = \sum_{i=1}^{N} \lambda_i \boldsymbol{u}_i \otimes \boldsymbol{v}_i$. Here, $\boldsymbol{u}_i$ and $\boldsymbol{v}_i$ are the $i$-th column vectors of the left and right singular matrices $\boldsymbol{U}$ and $\boldsymbol{V}$, respectively; $\lambda_i$ is the $i$-th singular value. The optimality of $\boldsymbol{u}_i \otimes \boldsymbol{v}_i$ renders $\lambda_i \boldsymbol{u}_i \otimes \boldsymbol{v}_i$ the $i$-th largest contribution to $d\mathbb{W}$, and $\sum_{i=1}^{m \ll N} \lambda_i \boldsymbol{u}_i \otimes \boldsymbol{v}_i$ the optimal $m$-dimensional reduced description. Because $d\mathbb{W} = \boldsymbol{F} \otimes d\boldsymbol{q}$, $\boldsymbol{u}_i$ and $\boldsymbol{v}_i$ are the optimal basis vectors for the force and displacement spaces, respectively.

The collection of all $\boldsymbol{u}_i$ forms an orthonormal coordinate system: $d\boldsymbol{s} = \boldsymbol{U}^T \cdot d\boldsymbol{q}$, termed the singular coordinates. $F_i = -\frac{\partial U}{\partial s_i}$ represents the force with the $i$-th largest impact on the system's dynamics, and $dW_i = F_i ds_i = \lambda_i (\boldsymbol{u}_i \cdot \boldsymbol{v}_i)$ is the $i$-th highest PEF in the system. Consequently, $\sum_{i=1}^{m \ll N} dW_i$ provides the optimal $m$-dimensional reduced description of $dU$, the master generating function of all EoMs in the system. This is the condition tRCs would conform to if they exist in a protein. Therefore, the singular coordinates provide the "ideal" coordinate system (Fig. 4); the leading singular coordinates should be tRCs, as they represent the directions of forces with the highest impact on the system's dynamics.

### 7.4 One-dimensional tRC in alanine dipeptide.

In ref. (72), Wu et al. applied the GWF method to the $C_{7eq} \rightarrow C_{7ax}$ transition of alanine dipeptide. The first singular coordinate, $R_c = 0.72\phi + 0.57\theta_1 + 0.34\tau_2 - 0.21\tau_1$, shows high PEF $\langle \Delta W_{R_c} \rangle = \langle \Delta W_\phi \rangle + \langle \Delta W_{\theta_1} \rangle$, while the other singular coordinates have zero PEFs. This demonstrates a cleaner energy flow pattern than in the internal coordinates (Fig. 4) and identifies $R_c$ as the one-dimensional tRC. In Fig. 2c, the accuracy of committor prediction increases from the intuition-based $\phi$ to the ML-based $\phi$ and $\theta_1$, and culminates in $R_c$ from the GWF method, demonstrating the superiority of the rigorous physics-based GWF method over empirical approaches.

**7.5 Multi-dimensional tRCs in HIV-1 protease.**

In ref. (133), Wu et al. applied the GWF method to the flap-opening process of HIV-1 protease (Fig. 5a) in an implicit solvent. They identified six tRCs that determine the committor with high accuracy (Fig. 5c). This marks the first successful identification of tRCs for a large-scale conformational change in a protein. Remarkably, the same procedure was used for both alanine dipeptide and HIV-1 protease without any system-specific information, demonstrating the GWF method's general applicability and effectiveness.

Further, applying bias potentials to the tRCs accelerates the flap opening by $10^4$-fold, confirming tRCs as the optimal CVs for enhanced sampling. Biasing along different tRCs results in multiple flap-opening pathways, suggesting a multi-dimensional physical picture. Each tRC is a linear combination of all the backbone dihedrals of the protease: $R_c = \sum_{i=1}^{N} c_i \chi_i$, highlighting the collective nature of protein conformational changes. The coefficients $c_i$ (Fig. 5b) define the cooperativity between different dihedrals $\chi_i$, illustrating how local protein movements, best described by individual $\chi_i$, accumulate into a large-scale conformational change. While collectivity in protein dynamics has long been recognized, its specific mechanism was unclear. The tRCs computed from the GWF precisely define this collectivity.

**7.6 Rigorous but economical test for validating tRCs.**

Committor test is critical for validating tRCs, but its high computational cost makes it impractical for large systems. Fortunately, results in ref. (133) suggest an economical but equally rigorous alternative. Figure 5d shows that biased trajectories along tRCs traverse intermediate committor values $p_B \in [0.1, 0.9]$, which mark the barrier transition period. Since conformations with such committor values can only be found along the natural transition pathway, Fig. 5d demonstrates that biased trajectories along the tRCs follow the natural transition pathway, despite their vastly accelerated time scale. This is a unique feature of tRCs unattainable by empirical CVs.

Consequently, a candidate RC can be validated as a tRC if it meets two conditions: 1) The target conformational change is substantially accelerated when bias is applied to the candidate RC, demonstrating effective enhanced sampling. 2) Biased trajectories along the candidate RC follow the natural transition pathway, as validated by using the shooting move (**sidebar callout: Shooting move**) of the TPS method to generate natural reactive trajectories (**sidebar callout: Natural reactive trajectory**) from these biased trajectories. Shooting move only succeeds when applied to TS conformations, proving that the biased trajectories pass through the TS. Since TS is the rarest conformation on the natural transition pathway, a biased trajectory passing through the TS most likely follows this pathway.

**7.7 tRCs from energy relaxation enable predictive enhanced sampling**.
A key advantage of tRCs in applications is that they provide optimal enhanced sampling. Previously, their practical use was limited because identifying tRCs requires natural reactive trajectories (62; 68; 71; 72; 133), which in turn need TS conformations that are only found on natural reactive trajectories and transition pathway, creating a deadlock (55; 141). An important discovery in ref. (134) transformed the situation: the tRCs for HIV-1 protease flap opening are identical to the leading singular coordinates of the GWF calculated from its energy relaxation, suggesting the latter as tRCs. Applying bias to the leading singular coordinates accelerated flap opening and ligand unbinding in explicit solvent—a process with an experimental lifetime of $8.9 \times 10^5$ seconds—to just 200 ps in MD simulations, confirming them as the optimal CVs for enhanced sampling.

**7.7.1 tRCs enable efficient generation of natural reactive trajectories**. Importantly, applying the shooting move to biased trajectories along the singular coordinates generates natural reactive trajectories with high efficiency, showing that the biased trajectories follow the natural transition pathway and validating the singular coordinates as tRCs. The natural trajectories offer distinct advantages: they accurately mirror physical reality and focus on the transition period while bypassing the prolonged waiting time, leading to high computational efficiency and low costs.

In contrast, methods like metadynamics, adaptive biasing force, and Gaussian accelerated dynamics do not preserve the dynamics of natural processes due to their use of bias potentials.

Markov state models and milestoning rely on using empirical CVs to cluster end points of MD trajectories and stitching them together to create long dynamic trajectories. Because CVs are different from tRCs, the resulting trajectories are disconnected in the subspace spanned by tRCs, representing distorted dynamics. Weighted ensemble sampling offers continuity in dynamics, but the computational cost is prohibitively high without knowledge of the tRCs.

**7.7.2 Predictive sampling**. The discovery above allows the GWF method to compute tRCs from energy relaxation, which enable optimal enhanced sampling of conformational changes. Since it requires only a single protein structure as input, this method has predictive capability, which was tested on the allostery of the PDZ2 domain. All PDZ domains have identical structures in both apo and holo states, yet allosteric effectors can substantially change their ligand binding affinities. This contradicts the conventional paradigm that effectors alter ligand affinity by exploiting the structural differences between apo and holo states. PDZ allostery has puzzled the field for 25 years, as highlighted in a recent review article titled 'Allostery Frustrates the Experimentalists' (142; 143). Li and Ma computed tRCs from the energy relaxation of holo PDZ2 and generated natural reactive trajectories of ligand unbinding (134). These trajectories revealed previously unrecognized synergistic conformational changes at both the allosteric and ligand binding sites, suggesting that effectors can alter ligand binding affinity by interfering with these conformational changes. This provides a straightforward mechanism for PDZ allostery.

**7.7.3 Generalized fluctuation-dissipation relation in proteins**. The findings above reveal a surprising reciprocity between an activated process (rare fluctuation) and energy relaxation (dissipation), best understood through the energy flows in proteins (71; 75; 76). During activation, energy flows from low-capacity channels (i.e., non-RCs) to high-capacity ones (i.e., tRCs), ensuring efficient energy delivery to the active site. During energy relaxation, energy flows in the opposite direction—from the active site to the tRCs, and then disperses among the non-RCs— through the same channels to maximize efficiency (132). If this holds true for more proteins, it could suggest a general mechanism for protein function and extend the fluctuation-dissipation relation to much higher energy ranges than currently understood.

**8. Closing Remarks**

Since Fischer's lock-and-key model in 1894 (144), the extraordinary features of protein function have fascinated scientists for over a century, leading to many exceptionally creative ideas for understanding the underlying mechanisms. However, the physical mechanisms of protein functions have existed since protein molecules first appeared and have remained unchanged. Thus, it is crucial to test our ideas objectively against this enduring physical reality. The committor test of tRCs provide an optimal approach for this purpose.

Pioneering work in the late 1990s highlighted a dilemma concerning tRCs: they are critically important for understanding protein functions but are exceedingly difficult to obtain. Consequently, intensive efforts have been made to develop systematic methods for identifying tRCs, beginning with the GNN method by Ma and Dinner. Meanwhile, many proteins have been well studied but none of them have their tRCs identified, limiting their mechanistic understanding.

The recent advancements of the energy flow theory and the GWF method have transformed the situation. Now, tRCs have been identified for both HIV-1 protease and the PDZ2 domain, revealing previously unattainable mechanistic insights. The ability to compute tRCs from energy relaxation further enables their identification at minimal computational cost. We expect this will open new avenues for investigating and understanding the correct mechanisms of protein functions.

Analysis of exact energy flows during the $C_{7eq} \rightarrow C_{7ax}$ transition of alanine dipeptide has revealed an intricate physical mechanism that markedly deviates from the standard Kramers picture. This finding calls for more systematic studies that could potentially enrich and refine the current theoretical framework of reaction dynamics and rates.

The optimality of tRCs for enhanced sampling is critical for free energy calculations, a key area in computational biophysics that involves two components: accelerating activated processes and subsequent unweighting or resampling. While the latter is well established, the former has been a bottleneck. tRCs provide the optimal solution to this challenge. Combined with tRCs' optimality for dimension reduction, this imparts great potential to energy flow theory and GWF for important applications such as allostery, ligand binding, enzymatic reactions, drug design, and enzyme design.

**Summary Points**
- tRCs—which meet the committor criterion—provide optimal enhanced sampling, optimal dimension reduction, and accurate reaction rate calculation.
- The GWF is a fundamental concept of Newtonian mechanics that transforms mechanical work—the generating function of EoMs and the core concept of mechanics—across different coordinate systems and generates the "ideal" coordinate system that cleanly separates tRCs from non-RCs.
- tRCs are distinguished by their high energy flows, underscoring their physical essence as the optimal energy flow channels in biomolecules.
- Both activation and relaxation in proteins are governed by the tRCs.
- tRCs in biomolecules are global and cooperative, enabling protein functions and distinguishing biomolecules from non-biomolecules.

**Sidebars or Terms and Definitions**

**Collective variables**: Collective variables (CVs) are empirical variables, typically constructed by intuition, for characterizing protein conformational changes. The primary requirement is that they can distinguish reactant from product states. Common examples of CVs include geometric parameters such as distance between protein domains, root mean square deviation from a reference structure, radius of gyration, and principal components. CVs are commonly used for enhanced sampling and dimension reduction. Some researchers also refer to CVs as reaction coordinates or order parameters, adding complexity to the terminology. For methods on finding CVs for general enhanced sampling and dimension reduction purposes, we refer readers to excellent reviews in the literature (145-148).

**Natural reactive trajectory**: A natural reactive trajectory is a regular MD trajectory that starts from the reactant state and ends in the product state. It covers the entire process of the system climbing and crossing the activation barrier during an activated process, while skipping the prolonged waiting time spent in the stable basins. Because the waiting time of an activated process is typically orders of magnitude longer than the time required for actually crossing the activation barrier, natural reactive trajectories provide a computationally efficient choice for studying activated processes. Since no bias is used in the simulation, a natural reactive trajectory faithfully simulates how an activated process occurs in reality. Therefore, it contains all the important information concerning the mechanism of an activated process.

**Natural transition pathway**: The natural transition pathway is the average pathway that a system follows during an activated process. It can be considered the center line of the ensemble of natural reactive trajectories and represents the minimum free energy pathway on the free energy landscape of the true reaction coordinates (tRCs).

**Transition path sampling**: Transition path sampling (TPS) is an efficient computational method for generating natural reactive trajectories. It requires an initial reactive trajectory as input, which serves as the seed for a Monte Carlo procedure in trajectory space. This procedure generates new reactive trajectories from existing ones, gradually covering the reactive trajectory space. The core of this Monte Carlo procedure is the shooting move, which is used for generating a new reactive trajectory from a conformation on an existing one.

**Shooting move**: In the shooting move, a conformation $R_0$ is selected from a dynamic trajectory, and momenta $p_0$ are drawn from the Boltzmann distribution. Two regular MD trajectories are then launched from $R_0$ with initial momenta $p_0$ and $-p_0$, respectively. If these two trajectories reach opposite basins, we leverage the time reversibility of classical mechanics to create a reactive trajectory. This is done by reversing the momenta along the trajectory ending in the reactant basin and merging the two trajectories at $R_0$. Since no bias is used, trajectories generated by the shooting move are natural reactive trajectories. The success rate of the shooting move is extremely low almost everywhere in the conformation space except for the small region near the transition state, as detailed by Hummer (141).

**Transition path time**: Transition path time is the interval between the last time point a reactive trajectory leaves the reactant state and the first time point it enters the product state.

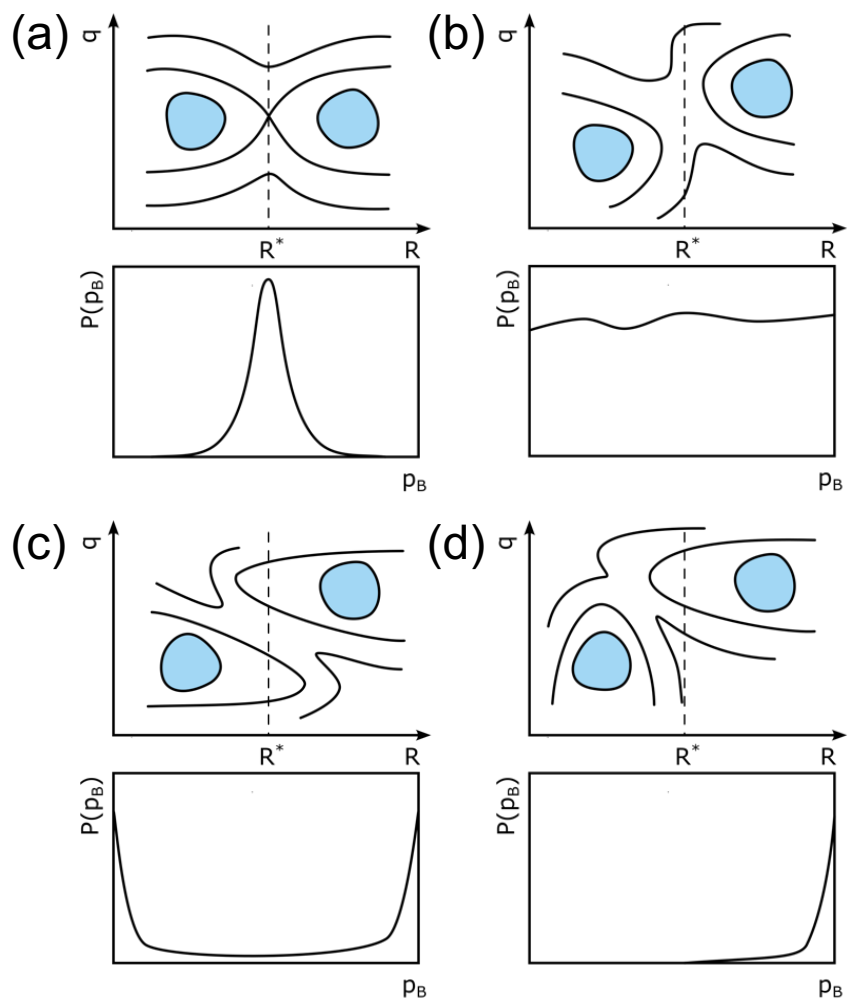

**Figure 1: Plausible outcomes of the committor test on a trial RC.**
Each panel shows a schematic two-dimension free energy landscape $F(r,q)$ (upper) and the corresponding committor distribution $P(p_B)$ (lower). Here, $R$ denotes the trial RC and $q$ a DoF orthogonal to it. These scenarios were first discussed in detail in the classical review article on the TPS method by Chandler and colleagues (55). **(a)** The tRC is $R$, thus $P(p_B)$ peaks at 0.5. **(b)** The motion along $q$, an important dynamic variable, is diffusive when $R$ is near $R^*$, resulting in a nearly constant $P(p_B)$. **(c)** The tRC has a significant component along $q$, resulting in a bimodal shape of $P(p_B)$. **(d)** The tRC is orthogonal to $R$, reflected by the single peak of $P(p_B)$ near $p_B = 0$. In this case, almost none of the conformations with $R = R^*$ are TS.

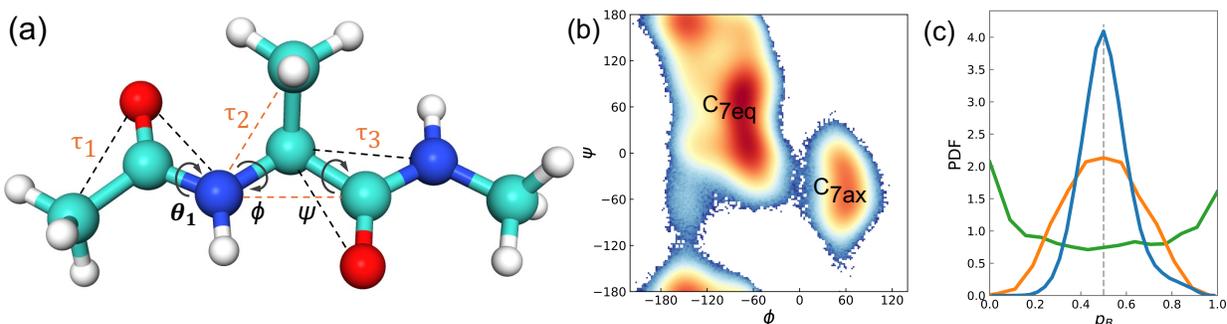

**Figure 2**: **Reaction coordinates for the $C_{7eq} \rightarrow C_{7ax}$ isomerization of alanine dipeptide in vacuum.**
**(a)** Molecular structure of alanine dipeptide. A curved arrow marks the bond of rotation for each dihedral. For an improper dihedral, the bond of rotation marks the edge shared by the two planes. Each plane is spanned by three atoms, with the central atom bonded to the other two. Two edges of each plane are marked by chemical bonds, while the third edge is marked by a dashed line connecting the two atoms that are not bonded to each other. **(b)** Definition of the $C_{7eq}$ and $C_{7ax}$ basins in the $(\phi, \psi)$-plane. The heat map is the logarithm of the joint probability $p(\phi, \psi)$ obtained from an equilibrium MD simulation. **(c)** Comparison of committor test results for different trial RCs. **Orange**: PDF of committor values of the ensemble of test conformations generated with $\phi$ as the trial RC. **Orange**: PDF of committor values of the ensemble of test conformations generated with both $\phi$ and $\theta_1$ as the trial RCs. **Blue**: committor values of the ensemble of test conformations generated with the 1D-RC identified by the GWF method.

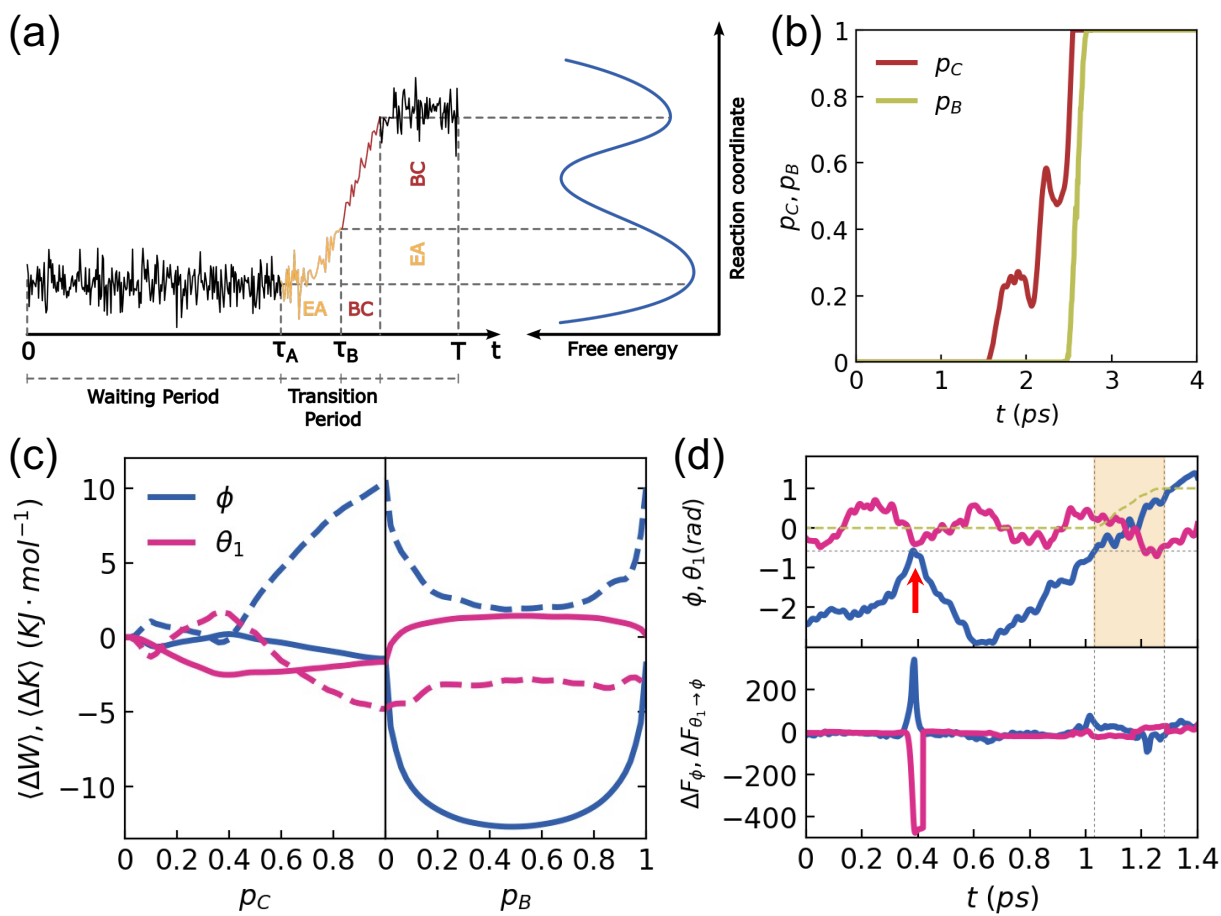

**Figure 3: Mechanism for the $C_{7eq} \to C_{7ax}$ transition of alanine dipeptide in vacuum from energy flow analyses.**
(a) A schematic showing different phases of an activated process in a double-well system. The noisy curve denotes a typical trajectory that eventually crosses the activation barrier and achieves the transition. The orange segment indicates EA and the red segment indicates BC. The time interval $[0, T]$ represent a typical reaction time and the duration of the trajectory. $\tau_A$ marks the onset of EA and $\tau_B$ marks the onset of BC. (b) Time evolution of the $p_C(\Gamma_0)$ and $p_B(R_0)$ along a typical reactive trajectory for the $C_{7eq} \to C_{7ax}$ transition. The reaction capacity $p_C(\Gamma_0) \in [0, 1]$ is the probability of finding a reactive trajectory in the vicinity of the phase space point $\Gamma_0$. It was developed by Wu and Ma to parameterize the EA phase (75), as committor is uniformly zero during EA. For the $C_{7eq} \to C_{7ax}$ transition, EA lasts much longer than BC. The end of EA is marked by $p_C = 1$, which coincides with the beginning of BC, marked by $p_B > 0$. (c) Ensemble averaged energy flows during EA (left panel) and BC (right panel). Solid lines denote PEFs, and dashed lines denote KEFs. The KEFs during BC are shifted so their values at $p_B = 0$ matches the KEFs at $p_C = 1$. Since $dW = -dU$, a negative PEF means an increase in potential energy. Both KEFs and PEFs are averaged over the reactive trajectory ensemble, using $p_C$ and $p_B$ as the projector for EA and BC phases, respectively. **(d)** This panel shows the cooperativity between $\theta_1$ and $\phi$ along a reactive trajectory. The upper panel shows the time evolution of $\phi$ (blue), $\theta_1$ (red), and committor (olive) along the reactive trajectory. The shaded region marks the BC period, where the committor changes from 0 to 1. This trajectory contains a failed attempt at BC (marked by the red arrow), where $\phi$ already reached the same value as the beginning of the successful BC. However, instead of proceeding to cross the activation barrier, it reverted to the $C_{7eq}$ basin because $\theta$ was not in the correct position. The lower panel shows that the force exerted on $\phi$ by $\theta$ (red curve) increases steeply in magnitude as $\phi$ tries to climb the barrier, counteracting the large force (blue curve) from all the other DoFs in the system that collectively push $\phi$ up the barrier, eventually pushing $\phi$ back. In contrast, $\theta$ is in the correct position during the successful BC. In this case, there was no large forces acting on $\phi$ and it crossed the activation barrier with ease. This result clearly demonstrates that proper movement of $\theta$ is necessary for $\phi$ to succeed in crossing the activation barrier.

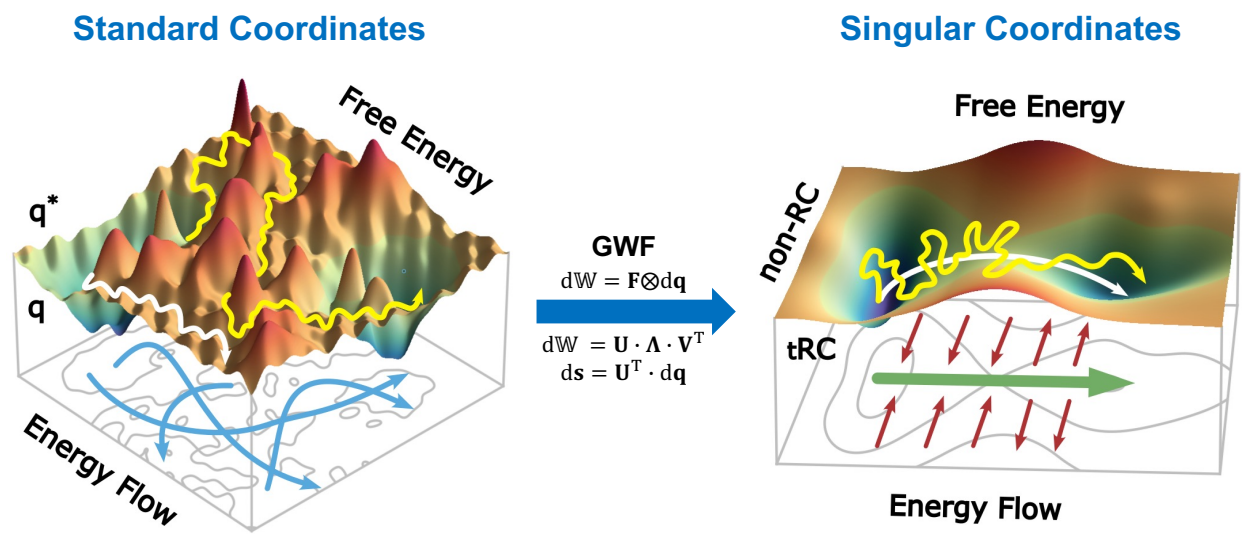

**Figure 4**: **A schematic of the physical picture of an activated process in standard versus 'ideal' coordinate system**.

The yellow curve on a free energy landscape represents a natural reactive trajectory, while the white curve represents a bias trajectory obtained by applying bias potential on $q$ and the tRC, respectively. The lower planes show energy flows within the system, with arrows indicating the direction of flow. When two curves cross each other, it indicates energy flow between them. In a standard coordinate system, the energy flows of different coordinates are similar in magnitude. In the 'ideal' coordinate system, energy flow through the tRC is much higher than through non-RCs. During activation, energy flows from non-RCs to the tRC. When the system relaxes into the stable basin, the energy flows in the opposite direction.

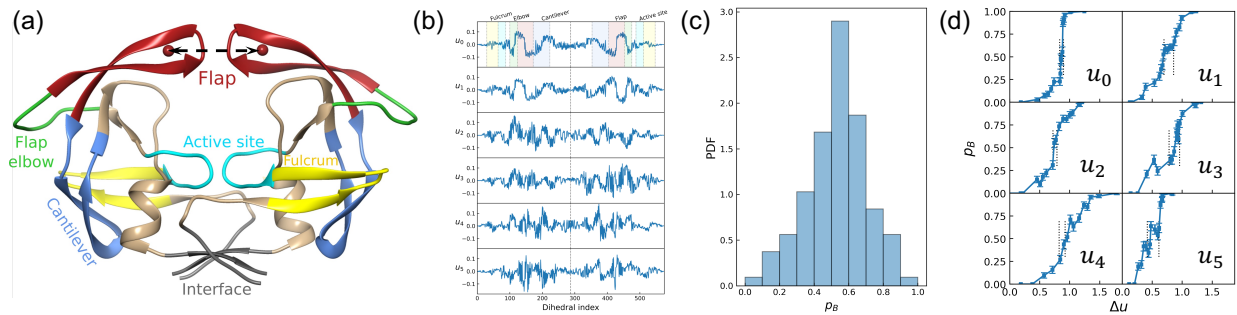

**Figure 5**: **Results on tRCs for HIV-1 protease flap opening in implicit solvent**.
**(a)** Crystal structure of the protease in semi-open state with different structural units labeled. **(b)** The coefficient vectors of the six tRCs. **(c)** Committor test results of the tRCs. **(d)** Committor values of conformations on biased trajectories along the tRCs.